\documentclass{llncs}
\usepackage{amssymb}
\usepackage{epsfig}
\usepackage{makeidx}  % allows for indexgeneration

\begin{document}

\begin{frontmatter}

\pagestyle{headings}  % switches on printing of running heads

\mainmatter

\title{Economic Models with Chaotic Money Exchange}

\titlerunning{Chaotic Money Exchange Models}  % abbreviated title (for running head)

\author{Carmen Pellicer-Lostao \and Ricardo L\'{o}pez-Ruiz}

\authorrunning{Pellicer-Lostao and L\'{o}pez-Ruiz  }   % abbreviated author list (for running head)

\institute{Department of Computer Science and BIFI, \\
Universidad de Zaragoza, 50009 - Zaragoza, Spain,\\
\email{carmen.pellicer@unizar.es,}
\email{rilopez@unizar.es} }

\maketitle              % typeset the title of the contribution

\begin{abstract}
This paper presents a novel study on gas-like models for economic systems. The interacting agents and the amount
of exchanged money at each trade are selected with different levels of randomness, from a purely random way to a
more chaotic one. Depending on the interaction rules, these statistical models can present different asymptotic
distributions of money in a community of individuals with a closed economy. {\bf Key Words:} Complex Systems,
Econophysics,  Gas-like Models, Money Dynamics, Random and Chaotic numbers, Modeling and Simulation
\end{abstract}

\end{frontmatter}

\section{Introduction}

Econophysics is born as a new science devoted to study Economy and Financial Markets
with the traditional methods of Statistical Physics ~\cite{mantegna}.
This discipline applies many-body techniques developed in statistical mechanics to
the understanding of self-organizing economic systems ~\cite{yakovenko2007}.
One of its main objectives is to provide economists with new tools and new insights to
deal with the complexity arising in economic systems.

The seminal contributions in this field ~\cite{yakovenko2000}, ~\cite{chacrabarti2000},
~\cite{bouchaud} have to do with agent-based modeling and simulation.
In these works, an ensemble of economic agents in a closed economy is interpreted
as a gas of interacting particles exchanging money instead of energy.
Despite randomness is an essential ingredient of these models,
they can reproduce the asymptotic distributions of money, wealth or income
found in real economic systems ~\cite{yakovenko2007}.

In the work presented here, the transfers between agents are not completely random as in the traditional
gas-like models. The authors introduce some degree of determinism, and study its influence on the asymptotic
wealth distribution in the ensemble of interacting individuals. As reality seems to be not purely random
~\cite{sanchez2007}, the rules of agent selection and money transfers are altered from random to pseudo-random
and extended up to chaotic conditions. This unveils their influence in the final wealth distribution in diverse
ways. This study records the asymptotic wealth distributions displayed by all these scenarios of simulation.

The paper is organized as follows: Section 2 introduces the basic theory of gas-like economic models. Section 3
describes the four simulation scenarios studied in this work, and the following sections show the results
obtained in the simulations. Conclusions are discussed in the final section.

\section{The gas-like model: Boltzmann-Gibbs distribution of money}

The conjecture of a kinetic theory of (ideal) gas-like model for trading in market was first discussed in 1995
~\cite{chacrabarti1995}. Then, it was in year 2000, when several noteworthy papers dealing with the distribution
of money and wealth ~\cite{yakovenko2000}, ~\cite{chacrabarti2000}, ~\cite{bouchaud} presented this theory in
more detail.

The gas-like model for the distribution of money assimilates the dynamics of a perfect gas, where particles
exchange energy at every collision, with the dynamics of an economic community, where individuals exchange money
at every trade. When both systems are closed and the magnitude of exchange is conserved, the expected
equilibrium distribution of these statistical systems may be the exponential Boltzmann-Gibbs distribution.

\begin{equation}
P(x) = a e^{-x/b}
\label{Eq.1}
\end{equation}

Here, $a$ and $b$ are constants related to the mean energy or money in the system,
$a=<x>^{-1}$ and $b=<x>$. Theoretically,
the derivation (and so the significance) of this distribution is based on the
statistical behavior of the system and on the conservation of the total magnitude of exchange.
It can be obtained from a maximum entropy condition ~\cite{jaynes1957} or from
purely geometric considerations on the equiprobability over all accessible states of the system ~\cite{ric2008}.

Different agent-based computer models of money transfer presenting
an asymptotic exponential wealth distribution can be found
in the literature ~\cite{yakovenko2000},~\cite{patriarca} ,~\cite{hayes}. In these simulations,
a community of $N$ agents with an initial quantity of money per agent, $m_0$, trade among them.
The system is closed, hence the total amount of money $M$ is a constant ($M=N*m_0$).
Then, a pair of agents is selected $(i,j)$ and a bit of money $\Delta m$
is transferred from one to the other. This process of exchange is repeated many times until
statistical equilibrium is reached and the final asymptotic distribution of money is obtained.

In these models, the rule of agents selection in each transaction is chosen to be random (no local preference
or no intelligent agents). The money exchange $\Delta m$ at each time is basically considered
under two possibilities : as a fixed or as a random quantity. From an economic point of view,
this means that agents are trading products at a fixed price or that prices (or products) can vary freely,
respectively.

These models have in common that generate a final stationary distribution that is well fitted by the
exponential function. Perhaps one would be tempted to affirm that this final distribution is universal
despite the different rules for the money exchange, but this is not the case as it can be seen
in ~\cite{yakovenko2000},~\cite{patriarca}.

\section{Simulation scenarios}

Real economic transactions are driven by some specific interest (or profit) between the different interacting
parts. Thus, on one hand, markets are not purely random. On the other hand, the everyday life shows us the
unpredictable component of real economy. Hence, we can sustain that the short-time dynamics of economic systems
evolves under deterministic forces and, in the long term, the recurrent crisis happening in these kind of
systems show us the inherent instability of them. Therefore, the prediction of the future situation of an
economic system resembles somehow to the weather prediction. We can conclude that determinism and
unpredictability, the two essential components of chaotic systems, take part in the evolution of economy and
financial markets.

Taking into account these evidences, the study of gas-like economic models where money exchange can have some
chaotic ingredient is an interesting possibility. In other words, one could consider an scenario where the
selection rules of agents and regulation of products prices in the market are less random and more chaotic.
Specifically, mechanisms for pseudo-random and chaotic number generation are considered in this paper.

In the computer simulations presented here, a community of $N$ agents is given with an initial equal quantity of
money, $m_0$, for each agent. The total amount of money, $M=N*m_0$, is conserved in time. For each transaction,
a pair of agents $(i,j)$ is selected, and an amount of money $\Delta m$ is transferred from one to the other.
The rules for money exchange will consider a variable $\upsilon$ in the interval $(0,1)$,
not necessarily random, in the following way:

\begin{itemize}
    \item \textbf{Rule 1}: the agents undergo an exchange of money, in a way that agent $i$ ends up with a
    $\upsilon$-dependent portion of the total of two agents money, ($\upsilon*(m_i+m_j)$),
    and agent $j$ takes the rest ($(1-\upsilon)*(m_i+m_j)$) ~\cite{patriarca}.
    \item \textbf{Rule 2}: an $\upsilon$-dependent portion of the average amount of the two agents money,
    $\Delta m=\upsilon*(m_i+m_j)/2$), is taken from $i$
    and given to $j$ ~\cite{yakovenko2000}. If $i$ doesn't have enough money, the transfer doesn't take place.
\end{itemize}

As there are two different simulation parameters involved in these gas-like models (the parameter for selecting the
agents involved in the exchanges and the parameter defining the economic transactions), four different
scenarios can be obtained depending on the random or chaotic election of these parameters.
These scenarios are considered in the following sections and are described as:

\begin{itemize}
\item \textbf{Scenario I}: random selection of agents with random money exchanges. \item \textbf{Scenario II}:
random selection of agents with chaotic money exchanges. \item \textbf{Scenario III}: chaotic selection of
agents with random money exchanges. \item \textbf{Scenario IV}: chaotic selection of agents with chaotic money
exchanges.
\end{itemize}

It is worthy to say at this point, that the words {\it random} and {\it pseudo-random} express a slight
difference in the statistical quality of randomness. The pseudo-random and chaotic numbers are obtained with two
chaotic pseudo-random bit generators, selected to this purpose. These are described in ~\cite{suneel} and
~\cite{ourlncs}, and are based in two 2D chaotic systems: the H\'{e}non Map and the Logistic Bimap. Interactive
animations of them can be seen in ~\cite{wolf_henon}.

The particular properties of these generators ~\cite{suneel}, ~\cite{ourlncs} make them suitable for the purpose
of this study. They are able to produce pseudo-random and chaotic patterns of numbers that can be used as
parameters of the simulations. Basically, these generators have two parts: the output of the chaotic maps is
used as input of a binary mixing block that randomizes the chaotic signals and generates the final random
numbers. Then, one one side, it is possible to take the exit of the chaotic blocks and produce chaotic sequences
of numbers. On the other side, they can generate a sequence of numbers with a gradual variation of randomness by
controlling a delay parameter $P$ taking part in the binary mixing block.

 This last feature is obtained by
varying the \emph{shift factor} $P$ ($P>1$) in a way that the lower its value, the worse is the random quality
of the numbers generated. Specifically, there is also a $P_{min}$ (around $80$) above which the properties of
the generator can be considered of random quality.

As an example to show this gradual variation of randomness, the generator in ~\cite{ourlncs} is used to produce
different binary sequences and initial conditions of $S2$ (further details ~\cite{ourlncs}). These bits are
transformed in integers of $32$ bits and transformed to floats dividing by the constant $MAXINT=4294967295$.

\begin{figure}[\h]
\centerline{\includegraphics[width=4.5cm]{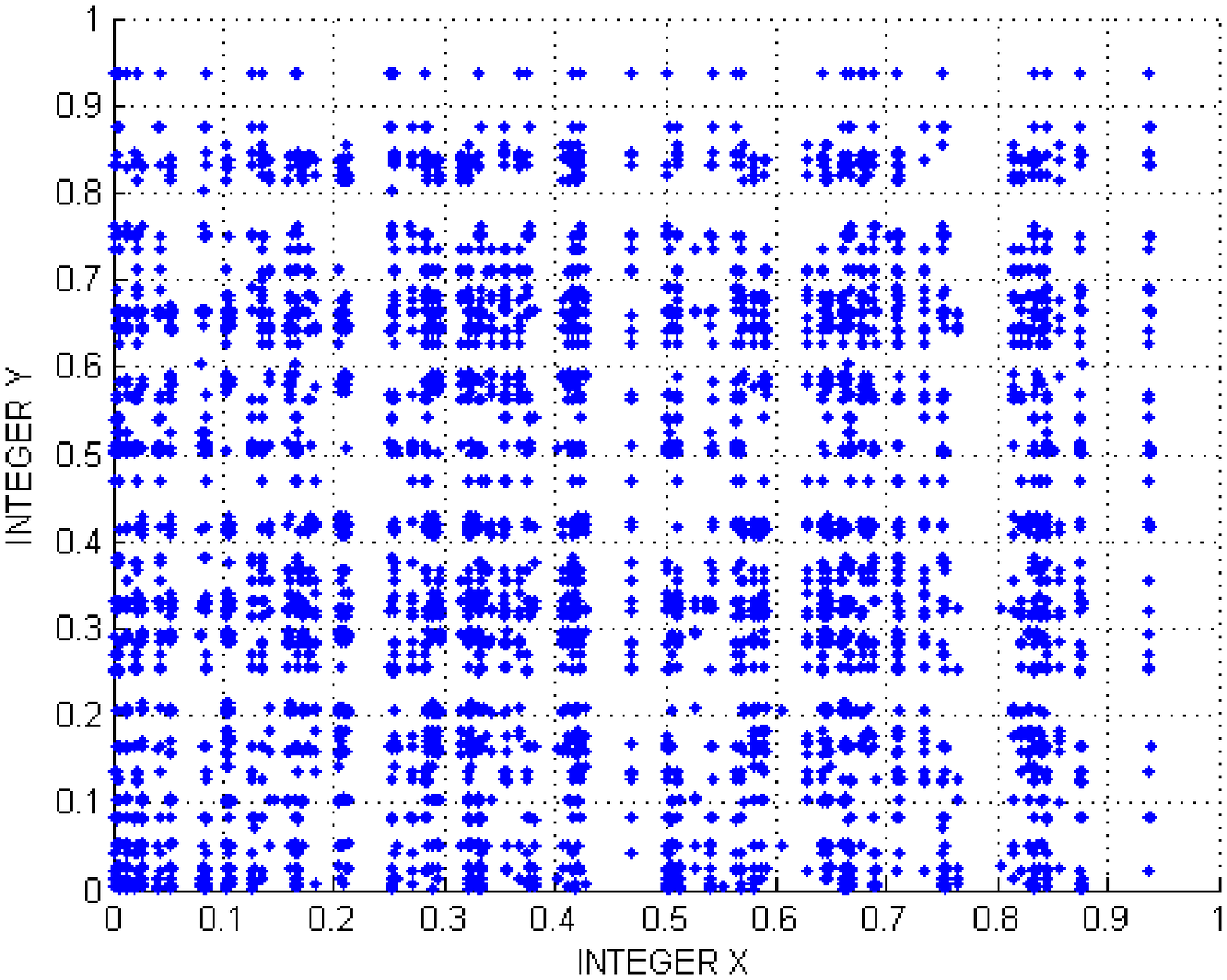}\hskip 1mm
\includegraphics[width=4.5cm]{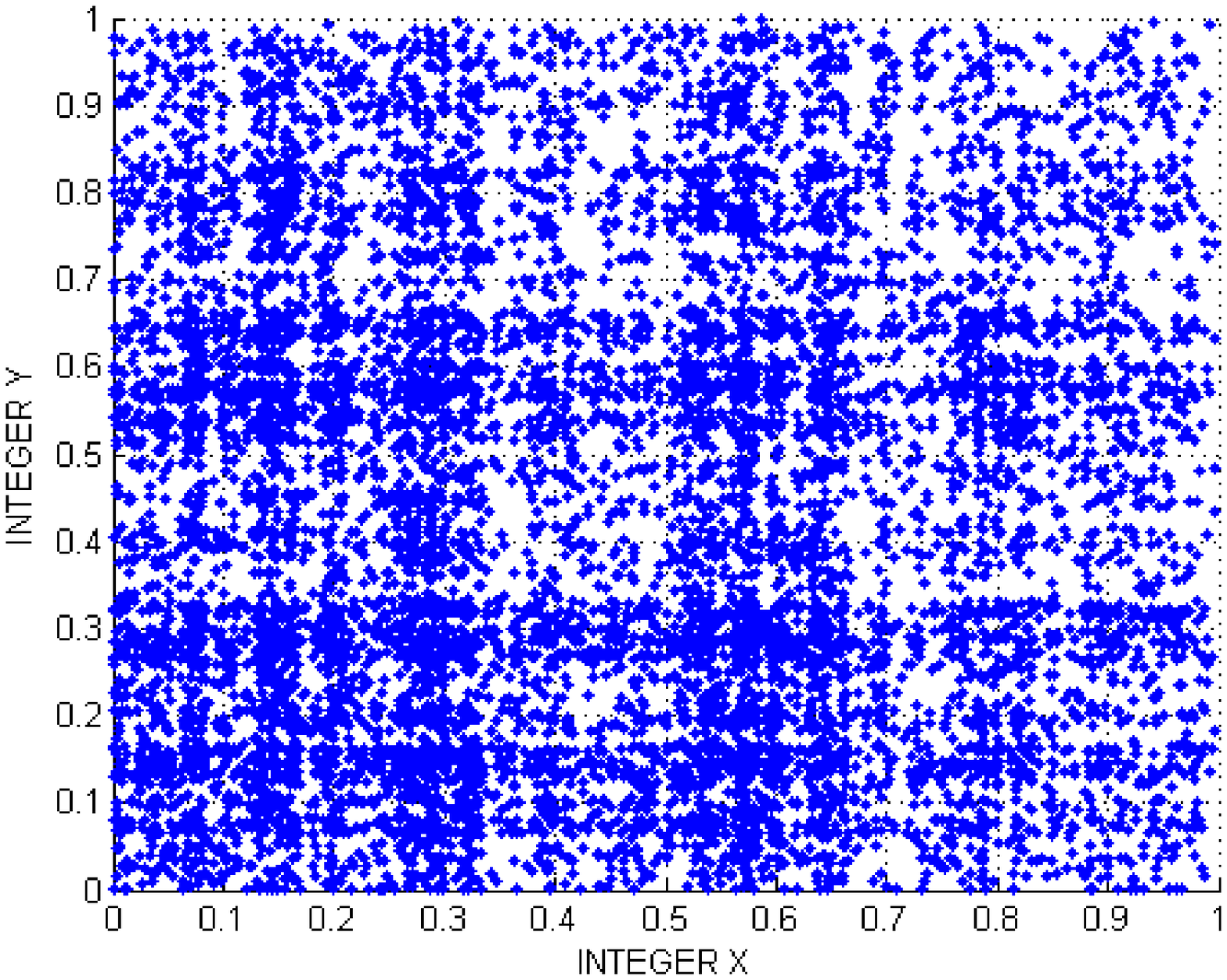}\hskip 1mm\includegraphics[width=4.5cm]{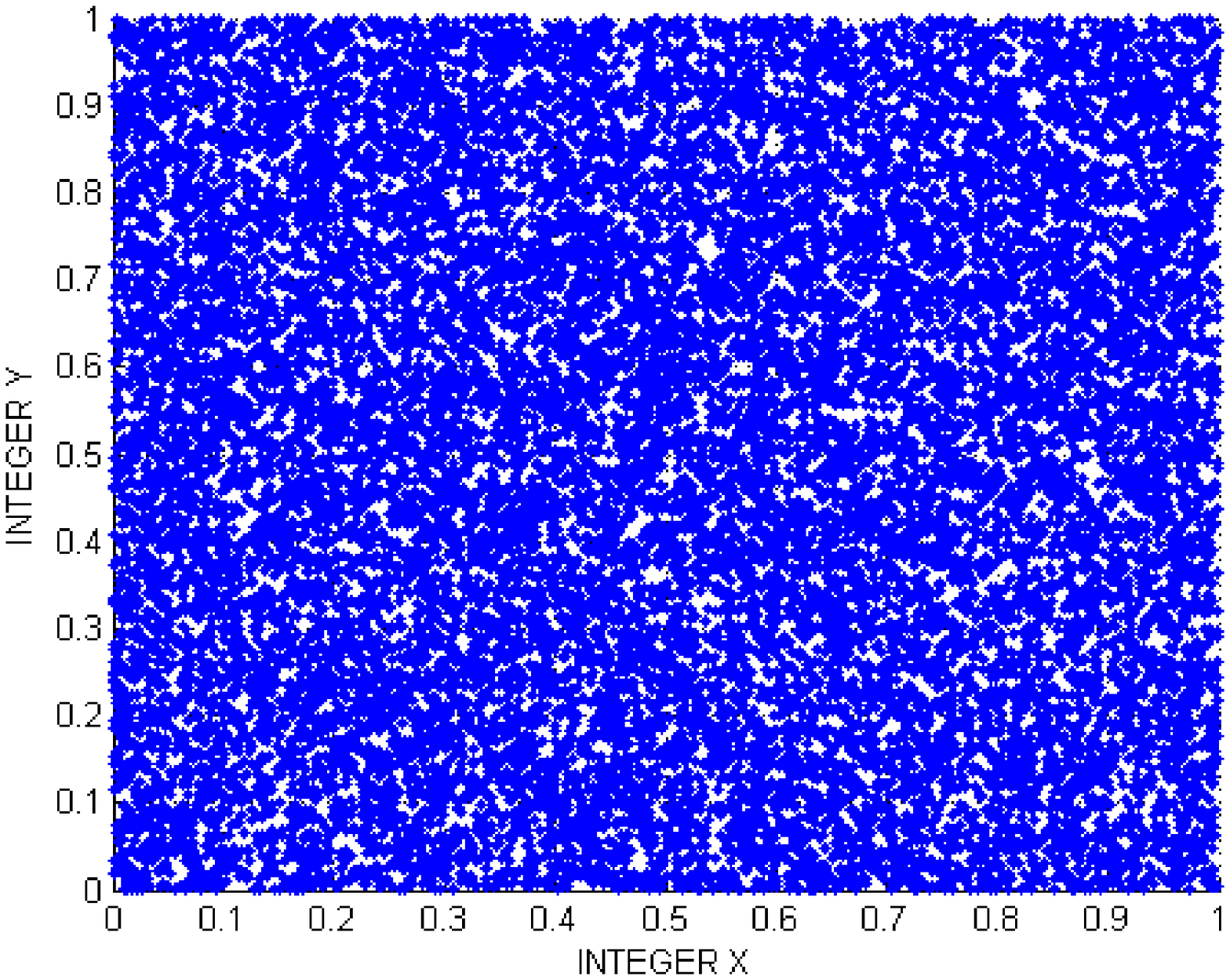}}
\centerline{(a)\hskip 4.5cm (b)\hskip 4cm (c)} \caption{Representation of $20000$ integers as pairs of floats in
the interval $[0,1]\times[0,1]$. The quality of their randomness improves as the shift factor $P$ grows in
magnitude. (a) $P=1$ (b) $P=5$ (c) $P=110$.} \label{fig2}
\end{figure}

When the shift factor $P$ is varied, the random quality of these binary sequences also varies. This can be
statistically measured by submitting them to statistical tests and it can be also graphically observed in
Fig.\ref{fig2}.

In Fig.\ref{fig2}, we generated $640000$ bits to obtain $20000$ integers of $32$ bits.
The integers obtained from the generator are transformed to flo
ats. The variation of the shift factor shows
graphically that with no shift at all, $P=1$, the integers obtained are hardly random. With $P=5$, the bits
generated do not pass the frequency or monobit test and still show a strong no random appearance. When the shift
factor grows over $P=110$, the binary sequences pass Diehard and NIST statistical tests. Graphically in
Fig.\ref{fig2}(d), it can be assessed to possess high random quality.

\section{Scenario I: Random selection of agents with random money exchange}

In this section, both simulation parameters are selected to obey certain pseudo-random patterns. Thus, the
generators in ~\cite{ourlncs} is used to produce different binary sequences (with initial conditions of $S2$ see
further details in ~\cite{ourlncs}). These bits are transformed in integers of $32$ bits and used as simulation
parameters to select the agents or the money to exchange.

Then, computer simulations are performed in the following manner.
A community of $N=500$ agents is considered with an initial quantity of money of $m_0=1000\$$. For each
transaction two integer numbers are selected from the generated pseudo-random sequence with a given shift factor
$P=P_{Ag}$. A pair of agents $(i,j)$ is selected according to these integers with an $N$-modulus operation.
Additionally, a third integer number is obtained from another pseudo-random sequence with another shift factor
$P=P_{Ex}$. This integer is used to obtained a float number $\upsilon$ in the interval $[0,1]$. The value of
$\upsilon$ and the rule selected (Rule 1 or Rule 2) for the exchange determine the amount
of money $\Delta m$ that is transferred from one agent to the other.

Choosing $P=P_{Ag}$ with different values it is possible to emulate an environment where the agents are locally
selected under a more or less random scenario situation. The same for $P=P_{Ex}$, the prices of products or
services in the market can be emulated to be less random, regulated, or completely arbitrary.

\begin{figure}[\h]
\centerline{
\includegraphics[width=4cm]{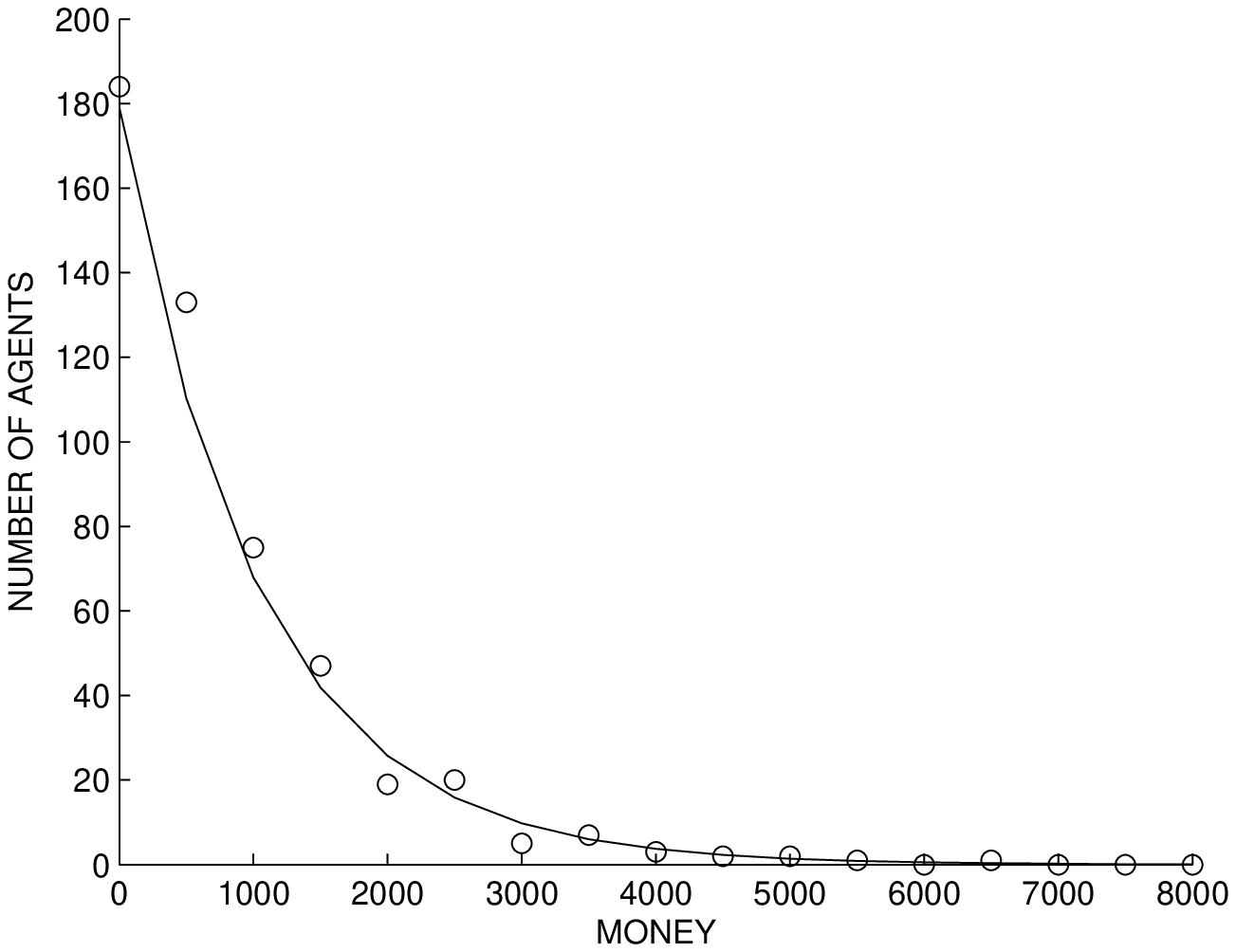}\hskip 1mm
\includegraphics[width=4cm]{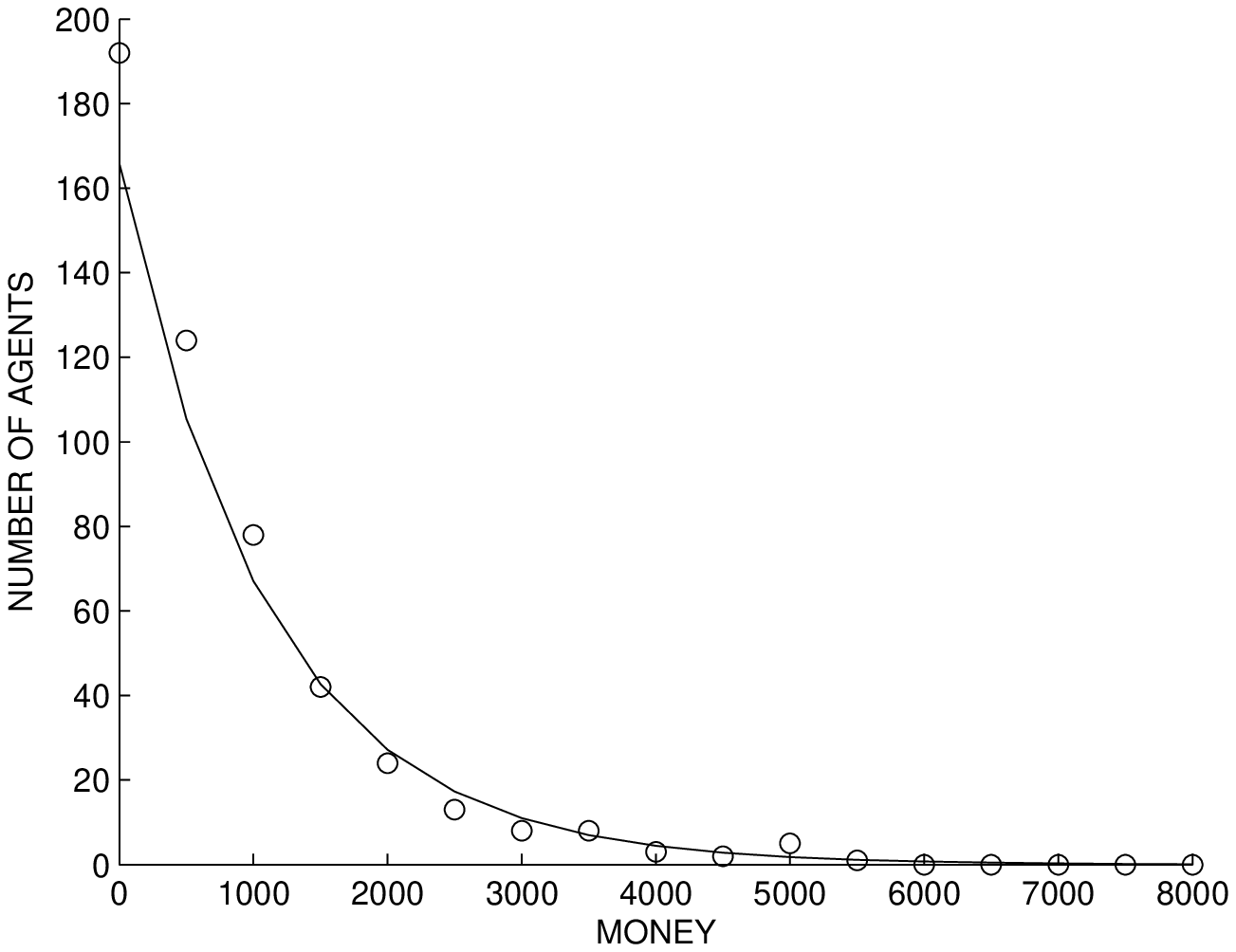}\hskip 1mm
\includegraphics[width=4cm]{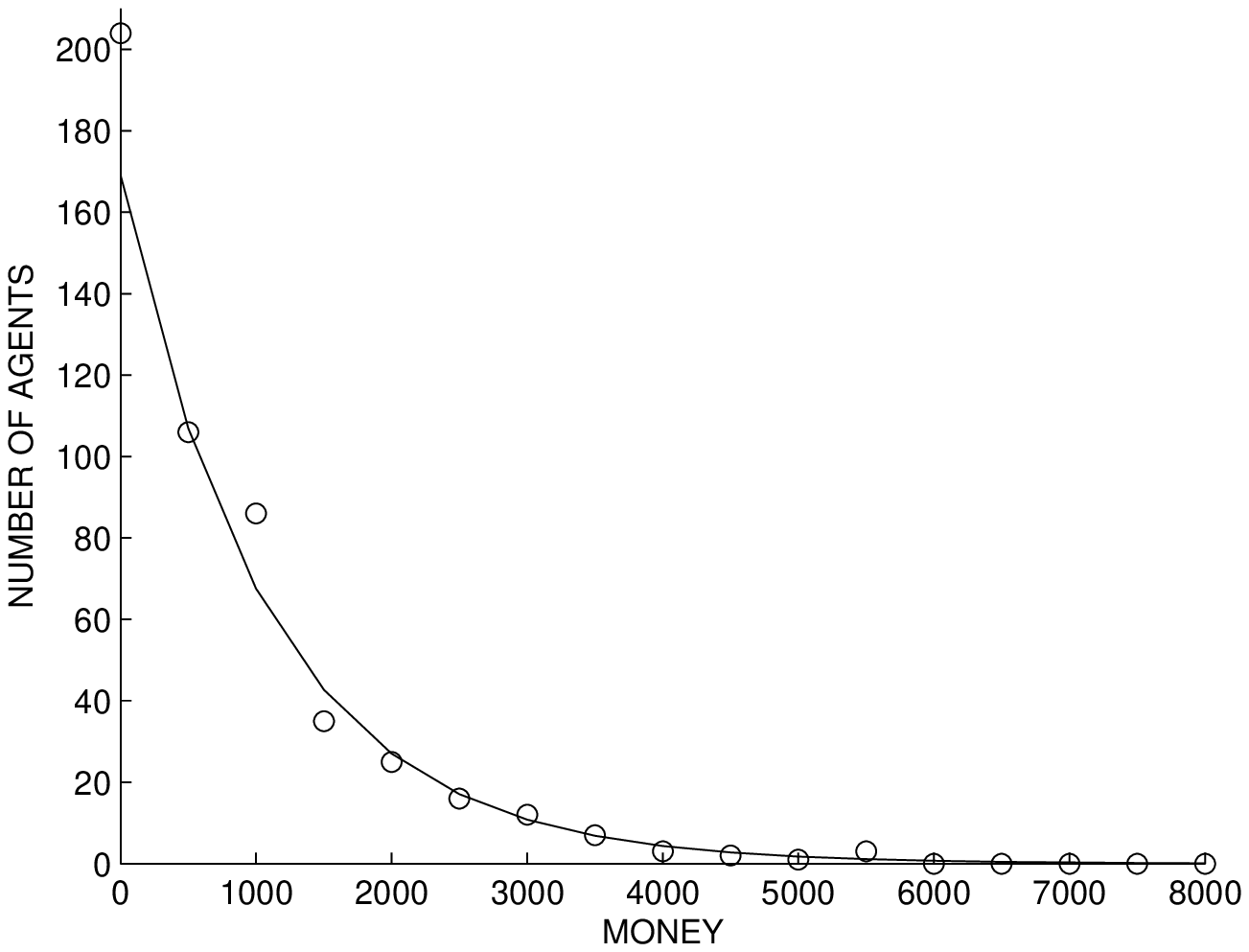}}
\centerline{(a)\hskip 4.5cm (b)\hskip 4cm (c)} \caption{Simulation of Scenario I. (a) $P_{Ag}=2$ and
$P_{Ex}=110$ for Rule 1, (b) $P_{Ag}=110$ and $P_{Ex}=3$ for Rule 2 and (c) $P_{Ag}=2$ and $P_{Ex}=5$ Rule 2.}
\label{fig3}
\end{figure}

The simulations take a total time of $400000$ transactions. In Fig. \ref{fig3}, different cases are considered,
taking pseudo random selection of agents or pseudo-random calculation of $\Delta m$. Two rules of money exchange
were considered, Rule 1 and 2 described in Section 2. The results show that all cases produce a stationary
distribution that is well fitted to the exponential function. Although not depicted in Fig. \ref{fig3} the case,
where both agents and traded money are selected randomly, gives very similar results to these cases of Fig.
\ref{fig3}, and also similar to the ones obtained in ~\cite{yakovenko2000}.

\section{Scenario II: Random selection of agents with chaotic money exchange}

In the previous section, it is observed that a variation in the random degree of selection of agents and/or
traded money, does not affect the final equilibrium distribution of money. It leads to an exponential in all
cases. In this section, the selection of agents is going to be set to random, while the exchange of money is
going to be forced to evolve according to chaotic patterns. Economically, this means that the exchange of money
has a deterministic component, although it varies chaotically. Put it in another way, the prices of products and
services are not completely random. On the other side, the interaction between agents is arbitrary and is chosen
randomly.

Taking the chaotic pseudo-random generators a step backwards, directly at the output of the chaotic block with
initial conditions $S2$ and $R1$ (see ~\cite{ourlncs} and ~\cite{suneel} respectively, for details), the chaotic
map variables $x_i$ and $y_i$ can be used as simulation parameters. Consequently, the computer simulations are
performed in the following manner. A community of $N=500$ agents is considered with an initial quantity of money
of $m_0=1000\$$. For each transaction two random numbers from a standard random generator are used to select a
pair of agents. Additionally, a chaotic float number is produced to obtain the float number $\upsilon$ in the
interval $[0,1]$. The value of $\upsilon$ is calculated as $|x_i|/1.5$ for the H\'enon map and as $x_i$ for the
Logistic Bimap. This value and the rule selected for the exchange determine the amount of money $\Delta m$ that
is transferred from one agent to the other.

\begin{figure}[\h]
\centerline{\includegraphics[width=4cm]{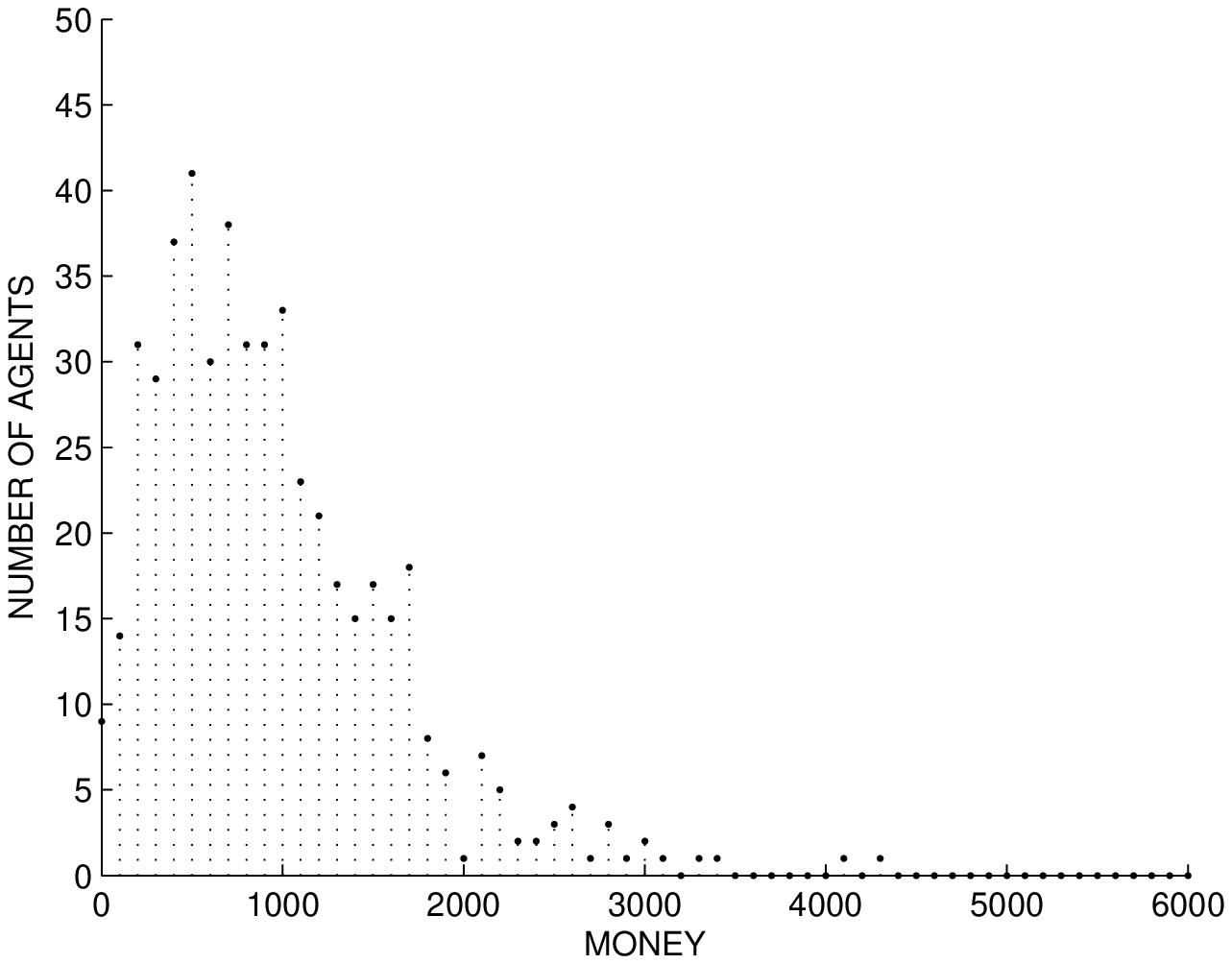}\hskip 1mm
\includegraphics[width=4cm]{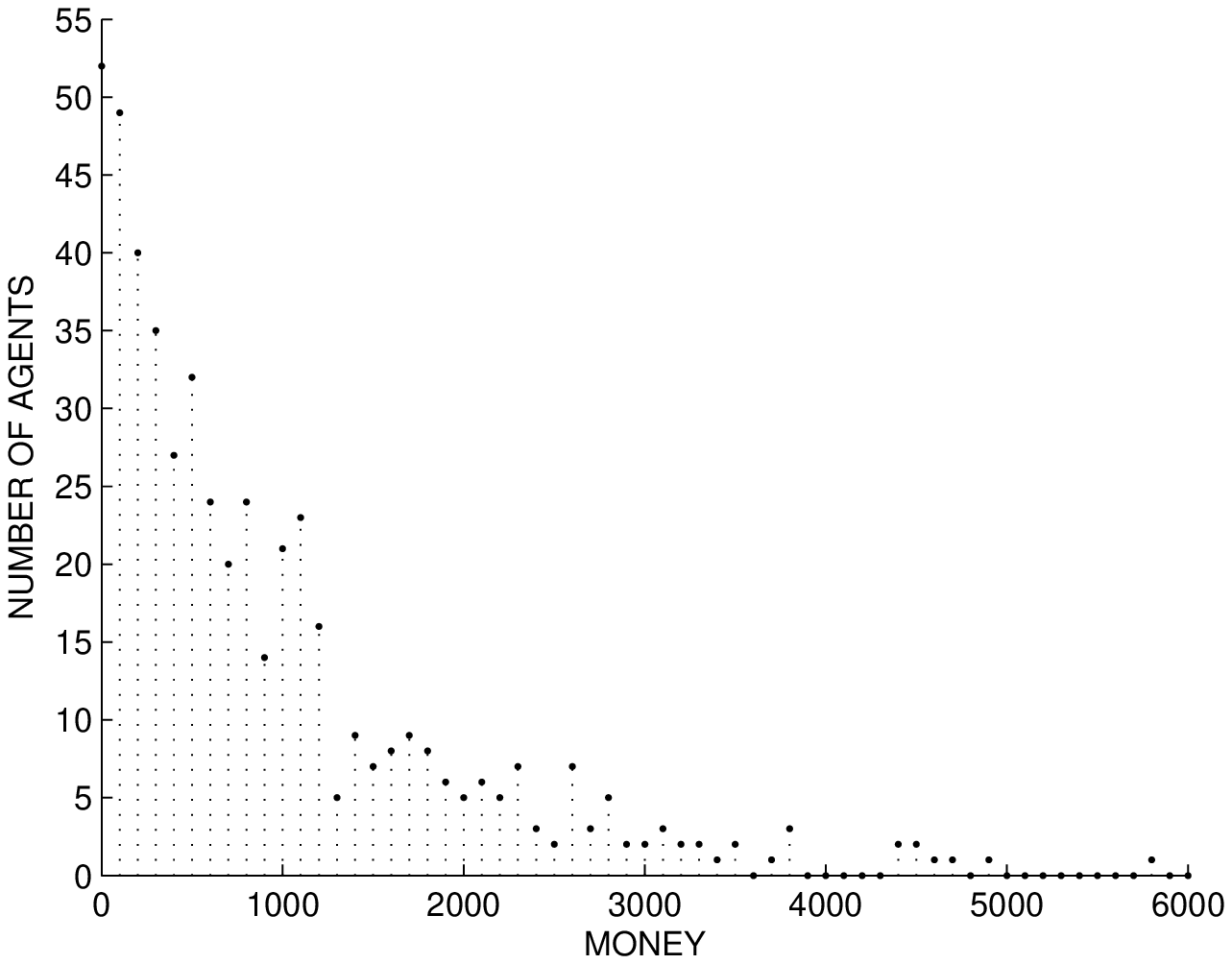}\includegraphics[width=4cm]{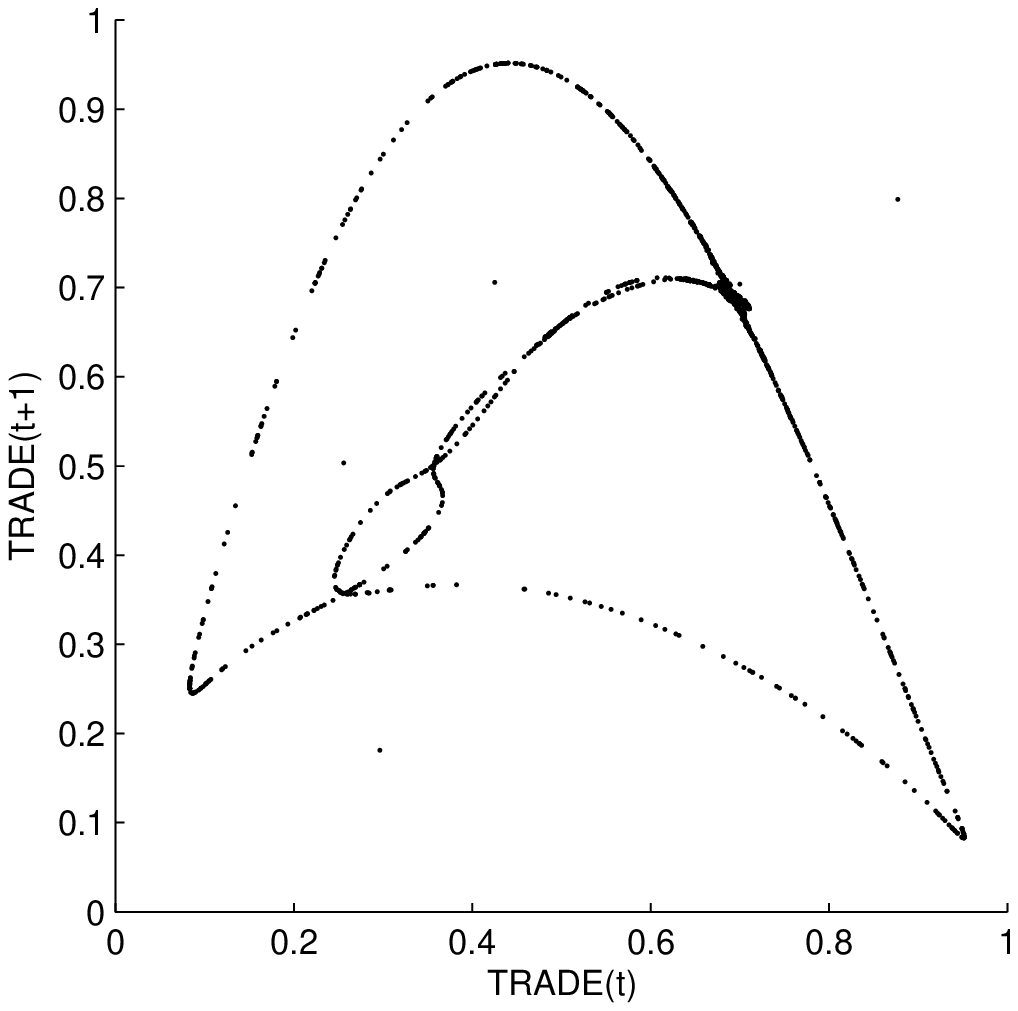}}
\centerline{(a)\hskip 4cm (b)\hskip 3.5cm (c)}
 \caption{Simulation of Scenario II where agents are selected
randomly and the money exchange follows chaotic patterns. (a) Chaotic trade selection with Logistic Bimap and
Rule 1, (b) Chaotic trade selection with Logistic Bimap and Rule 2 and (c) Chaotic trade points used in these
simulations represented as pairs in the range $[0,1]\times [0,1]$.} \label{fig4}
\end{figure}

The simulations take a total time of $400000$ transactions. Different cases are considered, taking the H\'enon
chaotic map or the Logistic Bimap. Rules 1 and 2 are also considered. New features appear in this scenario.
These can be observed in Fig.\ref{fig4}

The first feature is that the chaotic behavior of $\upsilon$ is producing a different final distribution for
each rule. Rule 2 is still displaying the exponential shape in the asymptotic distribution, but Rule 1 gives a
different function distribution. It presents a very low proportion of the population in the state of poorness,
and a high percentage of it in the middle of the richness scale, near to the value of the mean wealth. Rule 1
seems to lead to a more equitable distribution of wealth.

Basically, this is due to the fact that Rule 2 is asymmetric. Each transaction of Rule 2 represents an agent $i$
trying to buy a product to agent $j$ and consequently agent $i$ always ends with the same o less money. On the
contrary, Rule 1 is symmetric and in each interaction both agents $(i,j)$, as in a joint venture, end up with a
division of their total wealth. Now, think in the following situation: with a fixed $\upsilon$, let say
$\upsilon=0.5$, Rule 1 will end up with all agents having the same money as in the beginning, $m_0=1000\$$.
Using a chaotic evolution of $\upsilon$ means restricting its value to a defined region, that of the chaotic
attractor. Consequently, this is enlarging the distribution around the initial value of $1000\$$ but it does not
go to the exponential as in the random case ~\cite{patriarca}.

\section{Scenario III: Chaotic selection of agents with random money exchange}

In this section, the selection of agents is going to evolve chaotically, while the exchange of money is random.
Economically, this means that the locality of the agents or their preferences to exchange with each other is
somehow deterministic but with a complex evolution. Thus, some commercial relations are going to be restricted.
On the other hand, regulation of prices is random and they are going to evolve freely.

The chaotic generators are used directly at the output of the chaotic block, exactly as in the previous section.
Again a community of $N=500$ agents with initial money of $m_0=1000\$$ is taken and the chaotic map variables
$x_i$ and $y_i$ will be used as simulation parameters. For each transaction two chaotic floats in the interval
$[0,1]$ are produced. The value of these floats are $|x_i|/1.5$ and $|y_i|/0.4$ for the H\'enon map and $x_i$
and $y_i$ for the Logistic Bimap. These values are used to obtain $i$ and $j$ as in previous section.
Additionally, a random number from a standard random generator are used to obtain the float number $\upsilon$ in
the interval $[0,1]$. The value of $\upsilon$ and the selected rule determine the amount of money $\Delta m$
that is transferred from one agent to the other.

\begin{figure}[]
\centerline{\includegraphics[width=4cm]{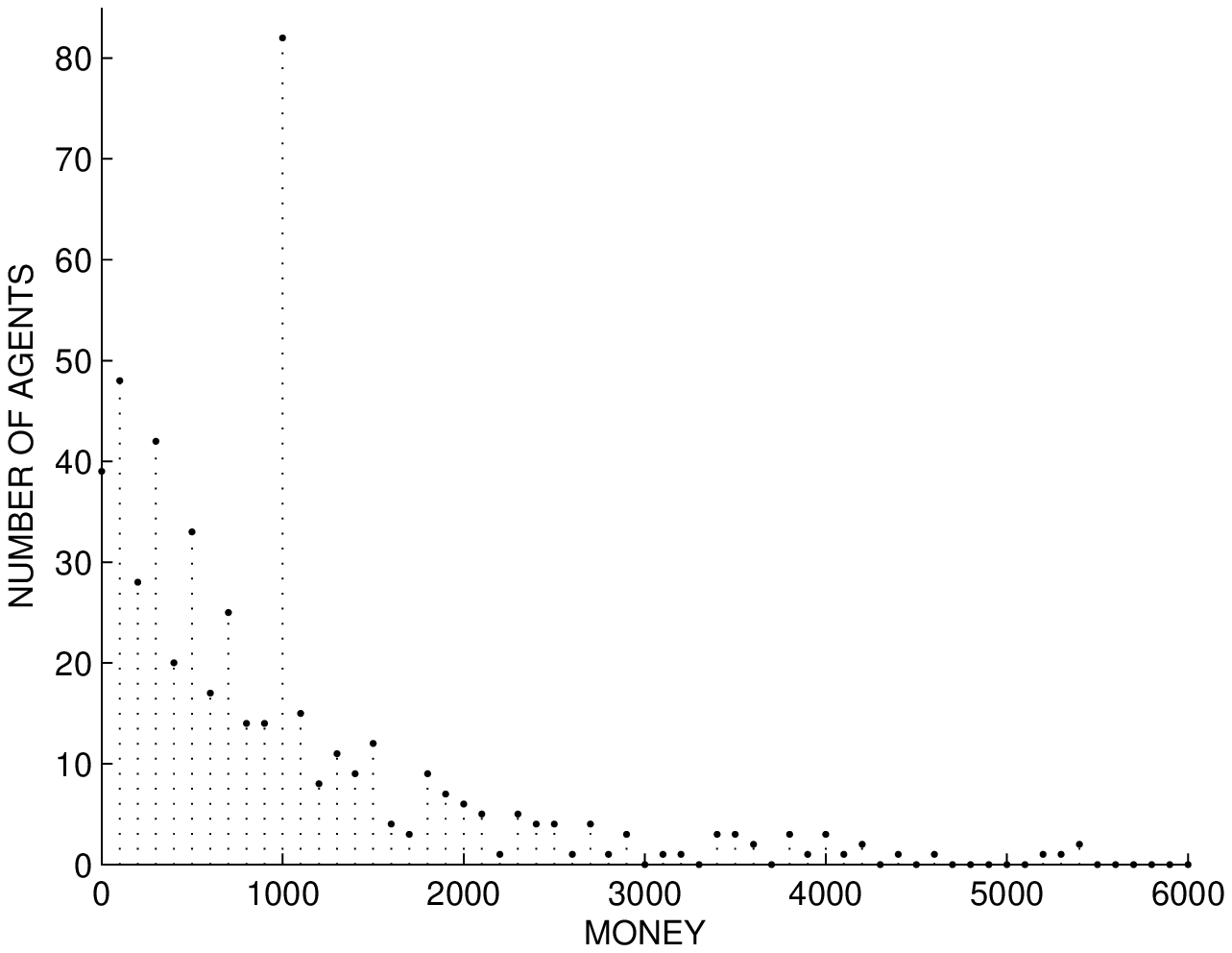}\hskip 1mm
\includegraphics[width=4cm]{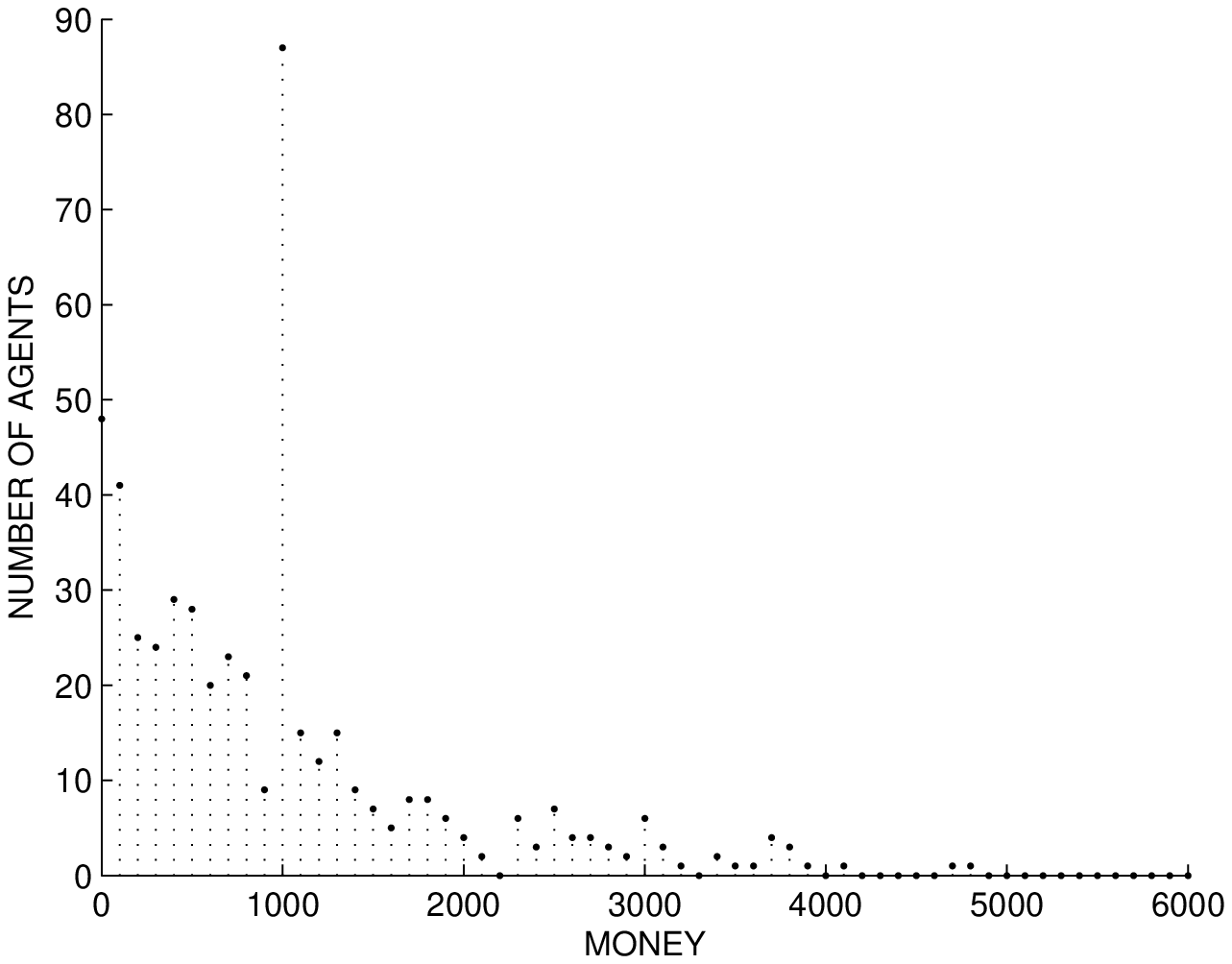}\includegraphics[width=4cm]{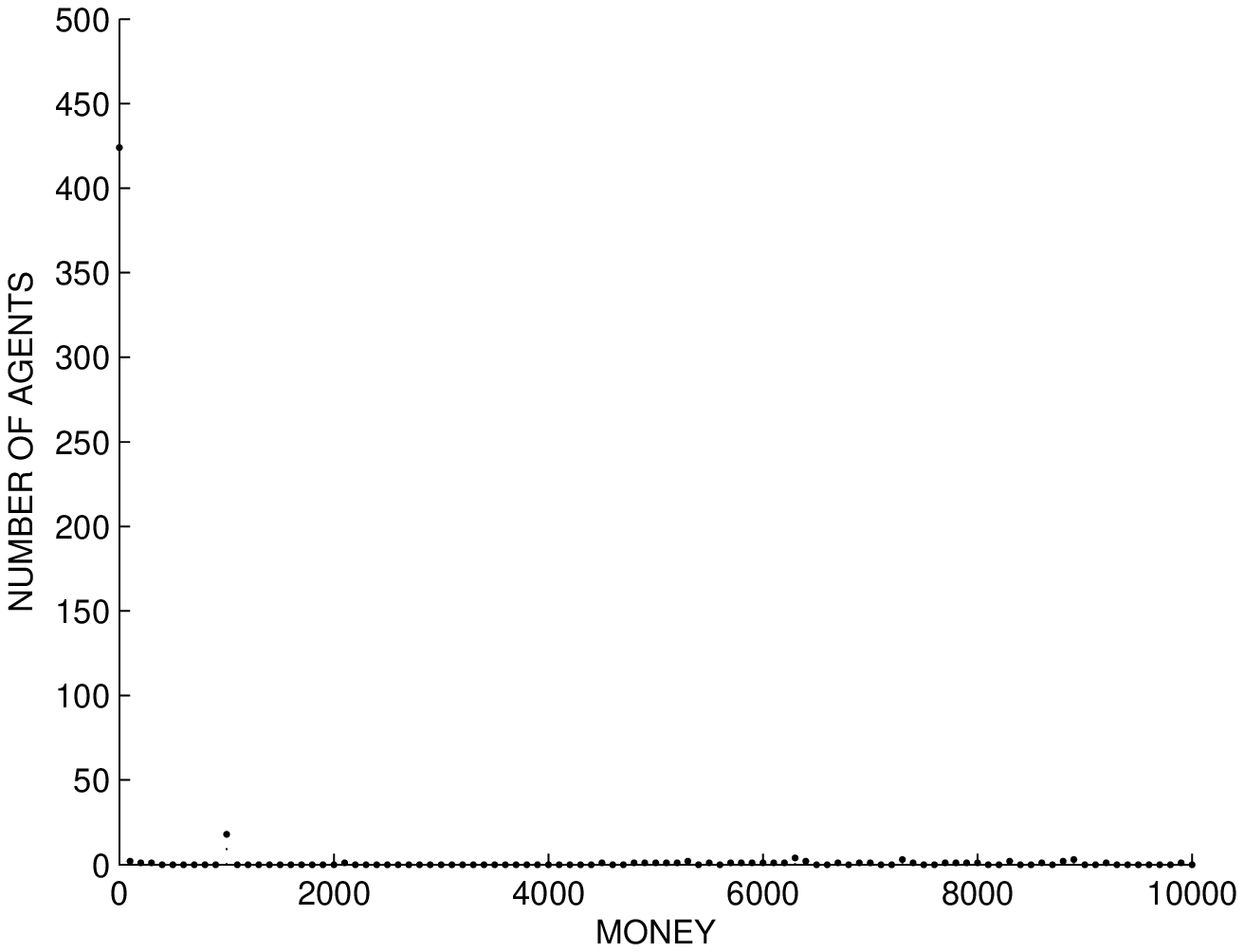}}
\centerline{(a)\hskip 4cm (b)\hskip 3.5cm (c)}
 \caption{Simulation of Scenario III where agents are selected
chaotically and the money exchange is set to be random. (a) Chaotic agents selection with Logistic Bimap and
Rule 1, (b) Chaotic agents selection with Logistic Bimap and Rule 2, (c) Heavy tail distribution of Chaotic
agents selection with H\'enon map and Rule 2.} \label{fig5}
\end{figure}

The simulations take a total time of $400000$ transactions. Different cases are considered, taking the H\'enon
chaotic map or the Logistic Bimap, and Rules 1 and 2. As a result, an interesting point appears in this scenario
with both rules. This is the high number of agents that keep their initial money in Fig. \ref{fig5}(a) and (b).
The reason is that they don't exchange money at all. The chaotic numbers used to choose the interacting agents
are forcing trades between a deterministic group of them and hence some commercial relations result restricted.

In can be observed in Fig. \ref{fig5}(a) and (b), that the asymptotic distributions in this scenario again
resemble the exponential function. The Logistic Bimap is symmetric (coordinates $x_i$ and $y_i$) and it produces
the effect of behaving like scenario II but with a restricted number of agents.

Amazingly, the H\'enon Map with Rule 2 leads to a distribution with a heavy tail, a Pareto like distribution. A
high proportion of the population (around 420 agents) finish in the state of poorness, while there are a
minority of agents with great fortunes distributed up to the range of $40000\$$. This is due to the asymmetry of
the rule, where agent $i$ always decrements its money, and the asymmetry of coordinates $x_i$ and $y_i$ in the
H\'enon chaotic Map used for the selection of agents. This double asymmetry makes some agents prone to loose in
the majority of the transactions, while a few others always win.

\section{Scenario IV: Chaotic selection of agents with chaotic money exchange}

In this section, the selection of agents and the exchange of money are chaotic. Economically, this means that
commercial relations are complex and some transactions are restricted. The exchange of money is going to vary
disorderly, but in a more deterministic way. The prices of products and services are not completely random.

As in the previous sections, the chaotic generators are used directly at the output of the chaotic block. Again
the chaotic map variables, $x_i$ and $y_i$, will be used as simulation parameters. The computer simulations are
performed in the following manner. A community of $N=500$ agents is considered with an initial quantity of money
of $m_0=1000\$$. For each transaction, four chaotic floats in the interval $[0,1]$ are produced. Two of these
floats are $|x_i|/1.5$ and $|x_{i+1}|/1.5$ for the H\'enon map or $x_i$ and $y_i$ for the Logistic Bimap. These
values are used to obtain $i$ and $j$ through simple multiplication (i.e.: $i=x_i*N+1$). Additionally, a chaotic
float number is produced to obtain the float number $\upsilon$ in the interval $[0,1]$. The value of $\upsilon$
is calculated as $(|y_i|+|y_{i+1}|)/0.8$ for the H\'enon map or as $(x_{i+1}+y_{i+1})/2$ for the Logistic Bimap.
This value and the selected rule of exchange determine the money $\Delta m$ that is transferred between agents.

\begin{figure}[\h]
\centerline{\includegraphics[width=4cm]{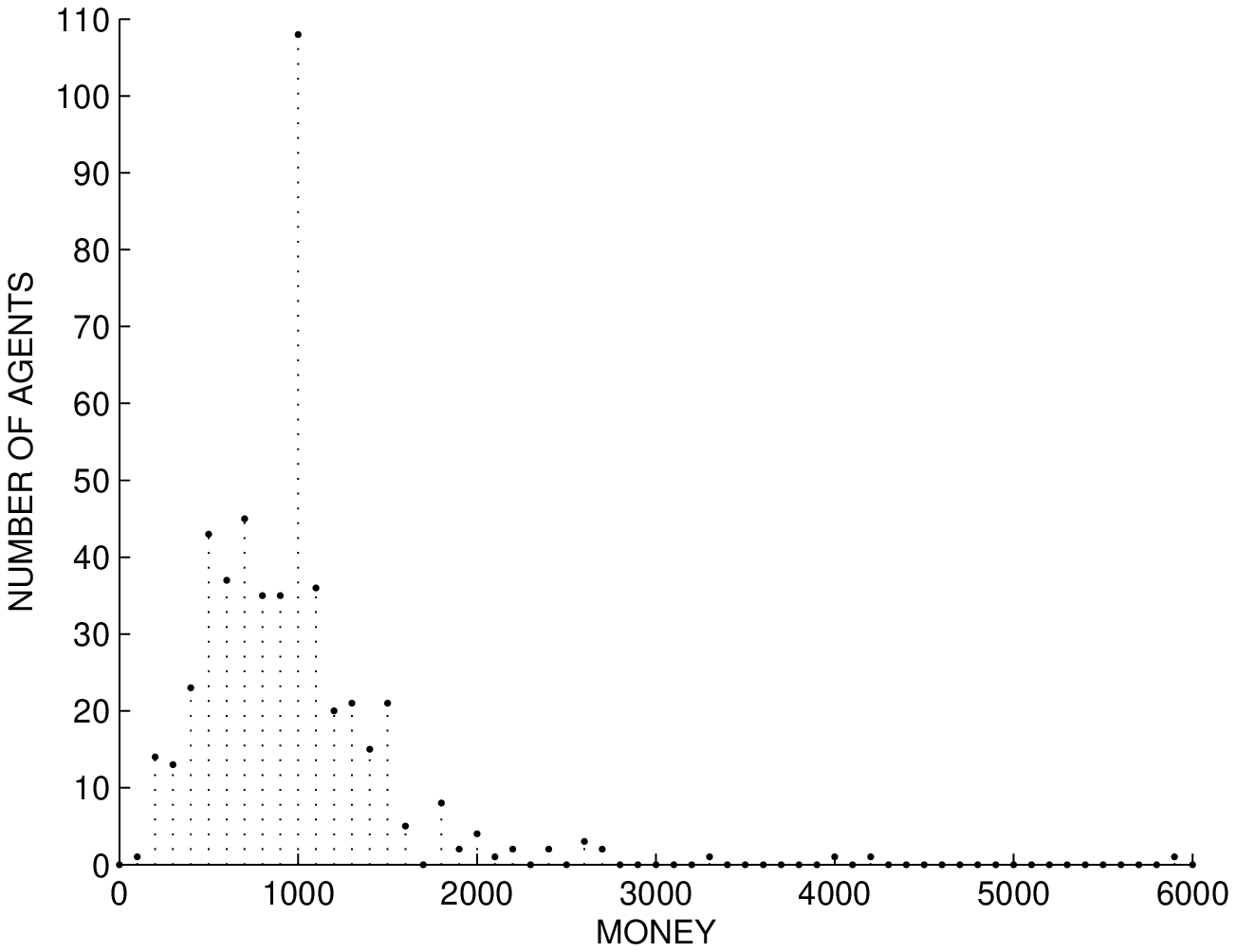}\hskip 1mm
\includegraphics[width=4cm]{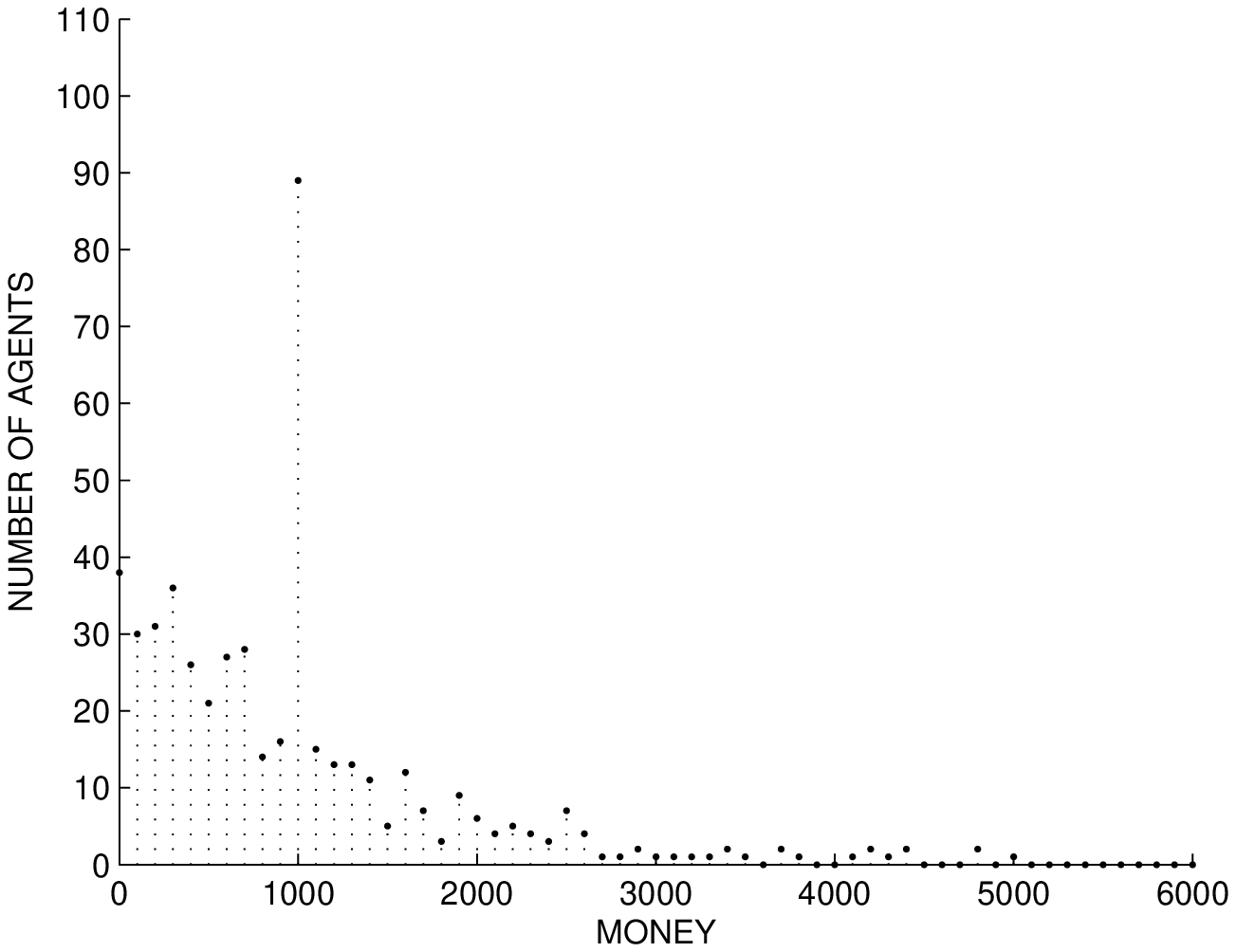}\includegraphics[width=4cm]{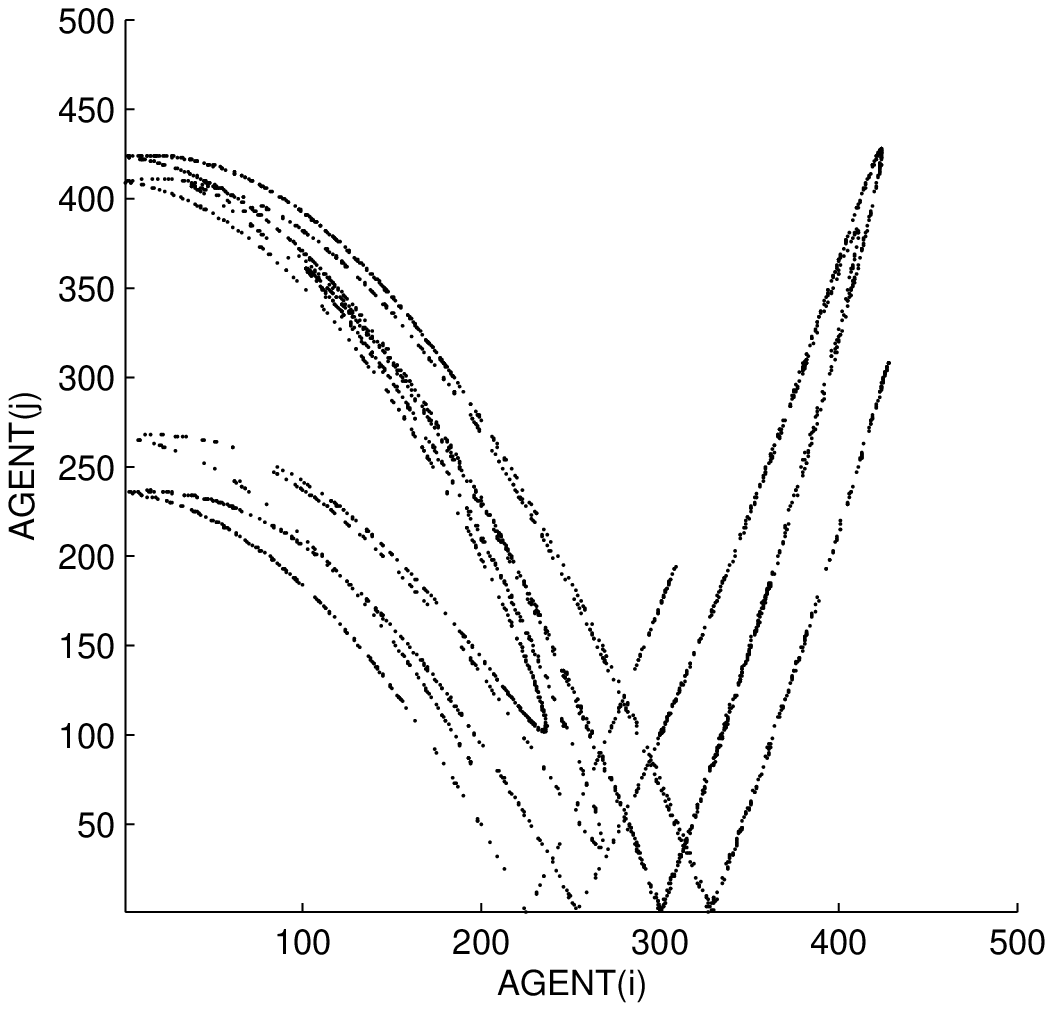}}
\centerline{(a)\hskip 4cm (b)\hskip 3.5cm (c)} \caption{Simulation of Scenario IV where agents and money
exchange are selected chaotically. (a) Chaotic selection with Henon Map and Rule 1, (b) Chaotic selection with
Henon Map and Rule 2 and (c) Chaotic agent points used in these simulations represented as pairs ($i$,$j$) in
the range $[1,500]\times [1,500]$.} \label{fig6}
\end{figure}

The simulations take a total time of $400000$ transactions. Also, in this scenario, we take the H\'enon chaotic
map or the Logistic Bimap, and Rules 1 and 2 are considered. As a result, the same properties of both asymptotic
distributions are maintained respect to section $5$, then the same differences between rules are observed,as
shown in Fig.\ref{fig6}

Here, again a high number of agents keep their initial money. The chaotic choice in Fig.\ref{fig6} (c) is
forcing trades between a specific group of agents, and then this type of locality makes some commercial
relations restricted. The different behavior for Rule 1 and 2 is similar to scenario III. Rule 2 still presents
an exponential shape, but Rule 1 gives a different function distribution with a maximum near the mean wealth. We
remark that Rule 1 is able to generate a more equitable society when chaotic mechanisms are implemented in both
processes, the agents selection and the money transfer.

\section{Conclusions}

The work presented here focuses on the statistical distribution of money in a community of individuals with a
closed economy, where agents exchange their money under certain evolution laws. The several theoretical models
and practical simulation results in this field, implement rules where the interacting agents or the money
exchange between them are traditionally selected as fixed or random parameters
(~\cite{yakovenko2007},~\cite{chacrabarti2000},~\cite{patriarca}).

Here, a novel perspective is introduced. As reality tends to be more complex than purely fixed or random, it
seems interesting to consider chaotic driving forces in the evolution of the economic community. Therefore, a
series of agent-based computational results has been presented, where the parameters of the simulations are
altered from random to pseudo-random and extended up to chaotic conditions.

In a first scenario, the exponential Boltzmann-Gibbs distribution is obtained for two different rules of money
exchange under pseudo-random conditions. Consequently, pseudo-randomness do not distinguish between different
evolution rules and richness is shared among agents in an exponential and unequal mode.

Introducing chaotic parameters in three other different scenarios leads to different results, in the sense that
restriction of commercial relations is observed, as well as a different asymptotic wealth distribution depending
on the rule of money exchange. It is remarkable that a more equitable distribution of wealth is obtained in one
of the evolution rules when some of the dynamical parameters are driven by a chaotic system. This can be
qualitatively observed in the distributions of money that have been obtained and reported for two different
scenarios.

The authors hope that this study may provide new clues in the nature of economic self-organizing systems.

{\bf Acknowledgements} The authors acknowledge some financial support by spanish grant
DGICYT-FIS200612781-C02-01.

%\newpage

\end{document}